\documentclass[fleqn,usenatbib]{mnras}
\pdfoutput=1
\usepackage{newtxtext,newtxmath}

\usepackage{graphicx}	
\usepackage{amsmath}	

\graphicspath{{img/}} 

\usepackage{times}
\usepackage{soul}
\usepackage{xspace}
\usepackage{color}
\usepackage[dvipsnames]{xcolor}
\usepackage{hyperref}

\def\snaw{SN\,2012aw\xspace}
\def\mdlA{\texttt{ R750M25Ni006E20}\xspace}
\newcommand\mdlB{\texttt{R500M25Ni006E20}\xspace}

\newcommand\nifsx{${}^{56}$Ni\xspace}
\newcommand\sneiip{SNe\,IIP\xspace}

\newcommand\stella{\textsc{STELLA}\xspace}

\newcommand{\editone}[1]{\textcolor{black}{#1}} 
\newcommand{\edittwo}[1]{\textcolor{black}{#1}}  
\newcommand{\editAn}[1]{\textcolor{black}{#1}}

\title[Parameters of SN 2012aw]{\editone{Parameters} of the type-IIP supernova SN\,2012aw}
\author[A.A.\,Nikiforova et al.]{
A.A.\,Nikiforova,$^{1,2}$\thanks{E-mail: a.a.nikiforova@spbu.ru}
P.V.\,Baklanov,$^{3,4}$
S.I.\,Blinnikov,$^{3,5}$
D.A.\,Blinov,$^{6,7,2}$
T.S.\,Grishina,$^{2}$
\newauthor
Yu.V.\,Troitskaya,$^{2}$
D.A.\,Morozova,$^{2}$
E.N.\,Kopatskaya,$^{2}$
E.G.\,Larionova$^{2}$
and I.S.\,Troitsky$^{2}$
\\
$^{1}$Pulkovo Observatory, St.-Petersburg, 196140, Russia\\
$^{2}$Astron. Inst., St.-Petersburg State Univ.,198504, Russia\\
$^{3}$NRC ``Kurchatov Institute'' -- Institute for Theoretical and Experimental Physics, Moscow 117218, Russia\\
$^4$National Research Nuclear University (MEPhI), Kashirskoe sh. 31, Moscow 115409, Russia\\
$^{5}$Dukhov Research Institute of Automatics (VNIIA), 127055, Moscow,
   Russia\\
$^{6}$Institute of Astrophysics, FORTH, Voutes,  Heraklion, 71110, Greece\\
$^{7}$Department of Physics, University of Crete, Heraklion, 71003, Greece\\
}

\date{Accepted XXX. Received YYY; in original form ZZZ}

\pubyear{2020}

\begin{document}
\label{firstpage}
\pagerange{\pageref{firstpage}--\pageref{lastpage}}
\maketitle


\begin{abstract}
We present the results the photometric observations of the Type IIP supernova SN 2012aw \editone{obtained for the time interval from 7 till 371 days after the explosion}.
\editone{Using the previously published values of the photospheric velocities we've computed the hydrodynamic model which simultaneously reproduced the photometry observations and velocity measurements}.
\editone{We found} the parameters of the pre-supernova: radius $R = 500 R_\odot$, nickel mass $M(^{56}$Ni$)$ $\sim 0.06 M_\odot$, \editone{pre-supernova mass} $25 M_\odot$, mass of ejected envelope $23.6 M_\odot$, explosion energy $E \sim \editAn{2  \times 10^{51}}$ erg.
\editone{The model progenitor mass $M=25 M_\odot$ significantly exceeds the upper limit mass $M=17 M_\odot$, obtained 
from analysis the pre-SNe observations}.
 \editone{This result confirms once more that the 'Red Supergiant Problem' must be resolved  by stellar evolution and supernova explosion theories in interaction with observations}.
\end{abstract}




\section{Introduction}
\label{intro}

Type IIP supernovae (\sneiip) are characterized by the presence of the ``plateau'' (region of almost constant luminosity) in the light curve, in contrast to types IIL and IIn, where the brightness decreases almost linearly after the maximum. Hydrogen lines and  P-Cygni profiles are observed in the spectra of \sneiip (as well as in the entire type II supernovae).

\editone{\sneiip are an important subject for research for a number of reasons. 
Supernovae play a critical role in the production and distribution of metals in galaxies, regulating star formation and galaxy evolution \citep{Nomoto2006}. 
The correlation between the parameters of the progenitor star and the observed parameters after a supernova explosion is not fully understood. The main factor why the slope changes during the plateau is not precisely defined, there are only assumptions \citep{Martinez2019}. 
\sneiip have been proposed as indicators of cosmological distances as an alternative to SNe Ia \citep{Hamuy2002}. 
There is a problem of progenitor masses, also known as "RSG problem",  which is that the mass estimated in hydrodynamic modeling ($15 - 25 M_\odot$) is usually more than the mass estimate taken from direct archived images of the progenitor ($9 - 17 M_\odot$) \citep{Smartt2009, Utrobin2009, Bersten2011, Smartt2009AA}. }

\editone{Hydrodynamic modeling of light curves is currently one of the most frequently used indirect methods for obtaining physical properties. We focused our attention on finding parameters using hydrodynamic modeling of one of the \sneiip, and also analyzed the results of calculations for this supernova that were published earlier.
}


We selected for research a bright supernova SN 2012aw. There is a quite detailed observational series for this supernova. We also present our observations in this article Section~\ref{observ}. Estimates of the parameters for the pre-supernova 2012aw were obtained in a number of works \citep{DallOra2014, Bose2013a, Martinez2019, Fraser2012, VanDyk2013} and in others. To calculate the model and determine the parameters of the pre-supernova, we used the \stella code \citep{Blinnikov2000, Blinnikov1998}. The details of our modeling are given in Section~\ref{Hyd_mod}.

 SN 2012aw was discovered March 16, 2012 by \cite{Fagotti2012} in the galaxy M~95 (NGC 3351). At that time, 
 \editone{its $R$ magnitude} reached $ R \approx 15^m$ \citep{DallOra2014}.
 \editone{We adopt an explosion epoch ($t_0$) of March 16.1, 2012 (JD=$2456002.6 \pm 0.8$ days) \cite{Bose2013a}.}

NGC 3351 is a SB(r)b spiral galaxy. 
There were no other supernovae detected in this galaxy before SN~2012aw. 

\editone{The distance estimates for NGC 3351 by different authors are rather  
similar. \cite{DallOra2014} adopt the distance}  modulus $29.96 \pm 0.04$~mag, as the average value obtained by two methods: with Cepheids and the top of the branch of red supergiants. 
In the work of \cite{Bose2013a} the distance was taken equal to $9.9 \pm 0.1$~Mpc, distance modulus
$29.97 \pm 0.03$. As in the previous case, the authors averaged the results from several assessment methods. \cite{Munari2013} took the distance modulus as $30.0 \pm 0.1$~mag, obtained by Cepheids. 
It can be seen that the values are very close to each other. 
For our calculations, \editone{we adopt} a distance modulus equal to $29.96\pm 0.04$~mag.

Total extinction from \cite{DallOra2014} was taken  $A(B)_{tot} = 0.36 \pm 0.07$ mag according to the excess color in our Galaxy $E(B - V) = 0.028$ mag, in the host galaxy $E(B - V) = 0.058 \pm 0.016 $ mag. 
In an article of \cite{Bose2013a} the total extinction was estimated as 
 $A_{v} = 0.23 \pm 0.03$ mag, the total color excess as $E(B - V) = 0.074 \pm 0.008$  mag. 
 \cite{VanDyk2013} \editone{estimate} total reddening as $E(B - V) = 0.077$  mag. 
 \cite{Fraser2012} obtained an estimate $E(B - V) = 0.10 \pm 0.05$ mag, 
noting that the value can be overestimated. 
\editone{We take the} total extinction value equal to $E(B - V) = 0.074 \pm 0.008$ mag \citep{Bose2013a}, since the result was obtained by averaging of several methods.

\section{OBSERVATIONS}
\label{observ}

\subsection{Observations and data reduction}

Observational data were obtained for the time interval \editone{from 7 till 371} days after the explosion. 

 The observations have been performed within the program of photometric and polaririmetric monitoring of variable sources carried in the Laboratory of Observational Astrophysics at St. Petersburg State University\footnote{\url{https://vo.astro.spbu.ru/en/node/17}}. The characteristics of the telescopes are presented in Table~\ref{obs}.

\begin{table*}
\caption{Characteristics of telescopes.}
\label{obs}
\begin{tabular}{c | c | c |}
\hline

Telescope    & AZT-8     & LX200\\
\hline
Diameter of the main mirror    &  700 mm         & 406 mm                \\

Focal length                   &  2780 mm        & 4060 mm               \\

Field of view                  & $8'.1 \times 5'.4$     & $14'.3 \times 9'.5$          \\

Optical scheme     & The main focus of the parabolic mirror &    Schmidt - Cassegrain  \\

CCD camera                     & ST-7 XME              & ST-7 XME  \\

Location & Crimean AO, P/O Nauchny, 600 m above sea level & AI, SPSU, 50 m above sea level  \\
\hline
\end{tabular}\\

\end{table*}

\begin{figure}
 \begin{center}
      \includegraphics[width=\columnwidth,clip]{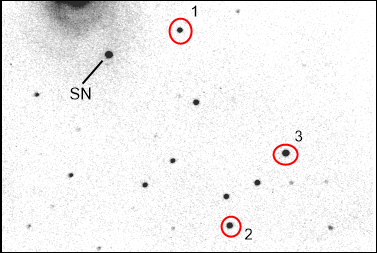}
      \caption{Supernova region of SN 2012aw. \editone{Circles indicate standard stars used for the photometry.}} 
 \label{figure1}
 \end{center}
 \end{figure}

The data have been processed using the standard utilities of  IRAF\footnote{\url{http://ast.noao.edu/data/software}}. The field stars used for the differential photometry are marked in Figure~\ref{figure1}. Their magnitudes listed in Table~\ref{mag} have been adopted from \cite{DallOra2014}.

\begin{figure}
\begin{center}
  \includegraphics[width=\columnwidth,clip]{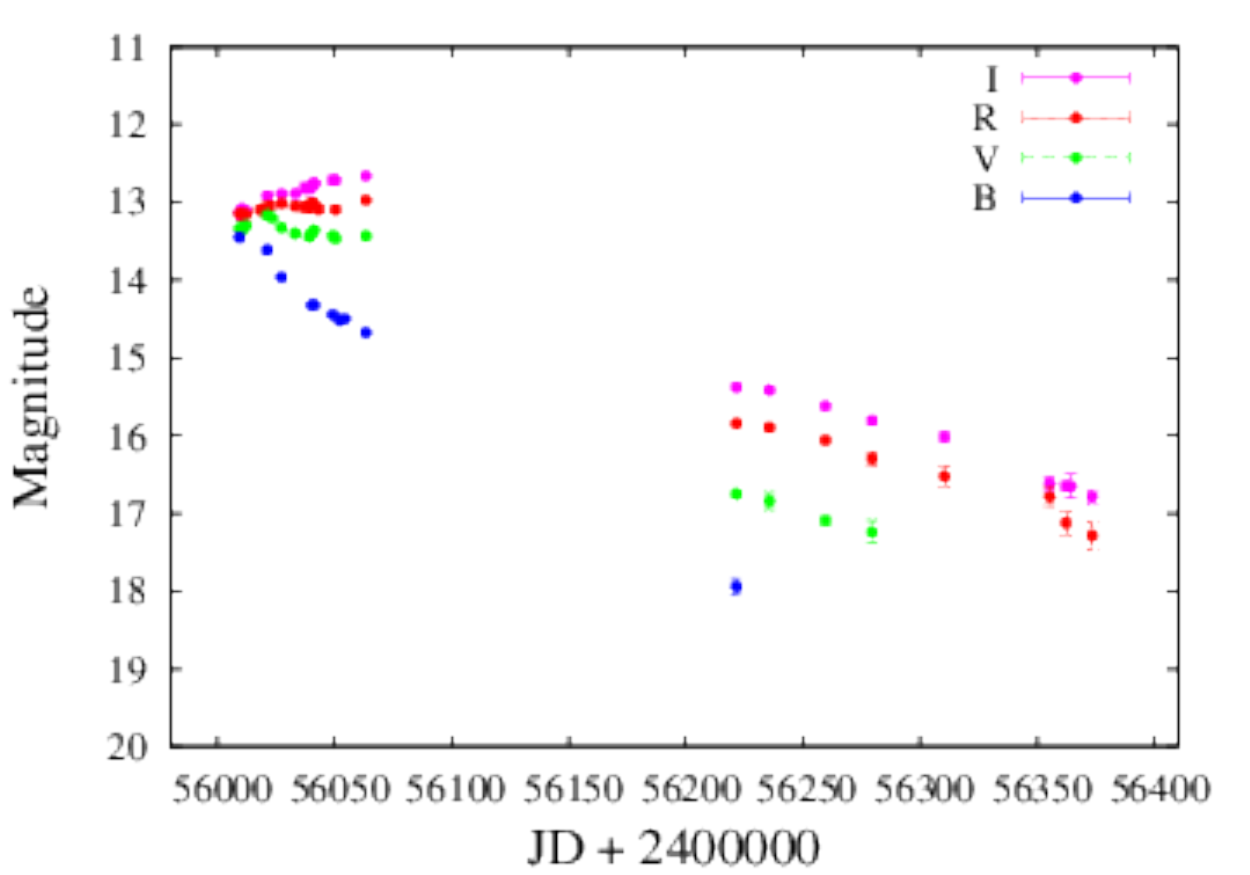}
   \caption{Light curve of \snaw according to the results of our photometry. Observations performed on telescopes AZT-8 and LX200 (see Table~\ref{obs}). The Y axis shows the apparent magnitude.}
\label{figure2}
\end{center} 
\end{figure}

\begin{figure}
\begin{center}
  \includegraphics[width=\columnwidth,clip]{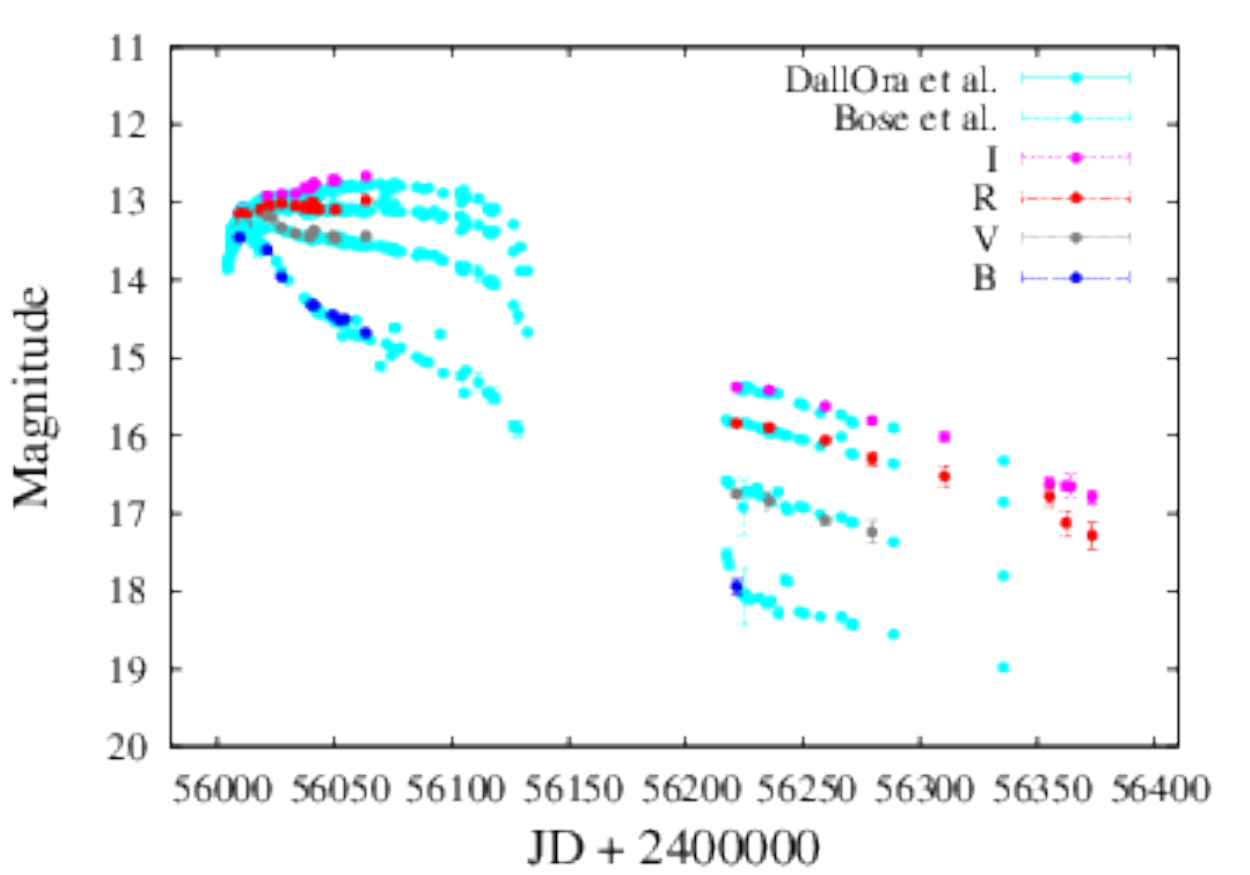}
   \caption{Comparison of the obtained light curve with photometric data from the literature \citep{DallOra2014, Bose2013a}. Blue dots indicate the results of photometry from the literature, other dots indicate the results of our photometry.}
\label{lc_all}
\end{center} 
\end{figure}

The light curves in four filters ($I$, $R$, $V$, $B$), obtained as a result of photometry, are shown in Figure~\ref{figure2}. A comparison of the results of our photometry
 with data from the literature \citep{DallOra2014, Bose2013a} is shown in Figure~\ref{lc_all}. The data are in quite good agreement with each other; our \editone{late time data} complementing declining part of the light curves.

\begin{table*}
\caption{Magnitudes of the standard stars marked in Figure~\ref{figure1}.}
\label{mag}
\begin{tabular}{c | c | c | c | c | c | c |}
\hline

 Star & $\alpha_{J2000}$ & $\delta_{J2000}$ & B (mag) & V (mag) & R (mag) & I (mag) \\
\hline
1 & $10^h43^m44^s.79$ & $+11^{\circ}41^{\prime}03^{\prime\prime}.84$ &15.351 & 14.972 & 14.706 & 14.450 \\
\hline
2 & $10^h43^m38^s.49$ & $+11^{\circ}35^{\prime}02^{\prime\prime}.17$ & 15.551 & 14.669 & 14.145 & 13.670 \\
\hline
3 & $10^h43^m31^s.42$ & $+11^{\circ}37^{\prime}16^{\prime\prime}.60$ & 14.992 & 13.932 & 13.248 & 12.717 \\
\hline
\end{tabular}

\end{table*}

\begin{figure}
\begin{center}
  \includegraphics[width=\columnwidth,clip]{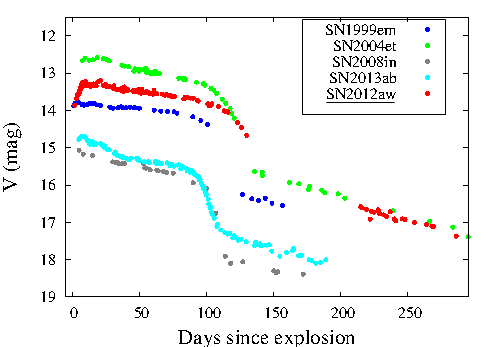}
   \caption{Comparison of the light curve of SN 2012aw with \sneiip: SN 1999em, SN 2004et, SN 2013ab and SN 2008in. The Y axis represents the apparent magnitude in filter $V$.}
\label{figure4}
\end{center} 
\end{figure}

\begin{figure*} 
 \begin{minipage}[h]{0.45\linewidth}
      \includegraphics[width=\columnwidth,clip]{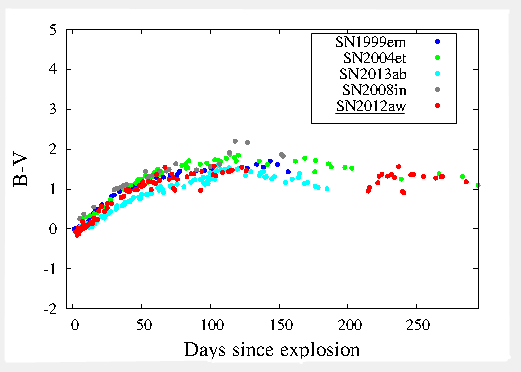}
      \end{minipage}
      \hfill
      \begin{minipage}[h]{0.45\linewidth}
      \includegraphics[width=\columnwidth,clip]{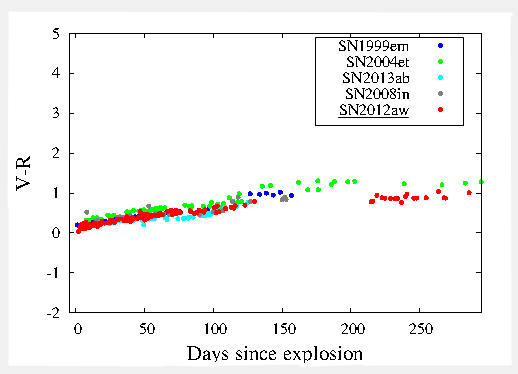}
      \end{minipage}
      \vfill
      \begin{minipage}[h]{0.45\linewidth}
      \includegraphics[width=\columnwidth,clip]{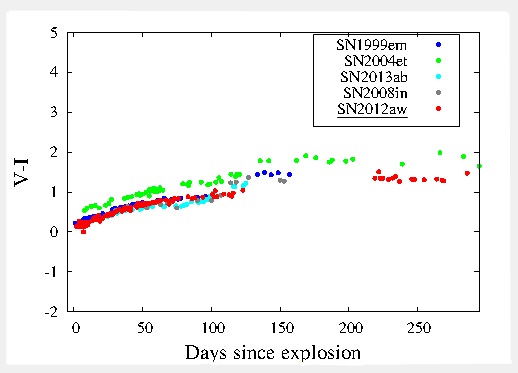}
      \end{minipage}
   \caption{Comparison of color indices of SN 2012aw and other \sneiip.}
   \label{figure5}
 \end{figure*}

 The Figure~\ref{figure4} shows a comparison of the light curve in $V$ band of supernova 2012aw with other \sneiip: SN 1999em, SN 2004et, SN 2013ab, SN 2008in \citep{Elmhamdi2003, Maguire2010, Bose2015, Roy2011}. It can be seen that the studied supernova fits well into the general picture of the  light curves of its type: there is a long region of the "plateau", followed by a sharp decline in brightness, which then goes on to the smooth and longest  phase of the ``tail''. Good agreement with other supernovae of that type can also be seen when comparing the SN 2012aw color indices with other \sneiip (Figure~\ref{figure5}).

\begin{table*}
\caption{Results of photometry of SN 2012aw obtained from AZT-8 and LX200 telescopes. The date $t_0$ is accepted as  $JD =  2456002.5 $.}

\label{table3}
\begin{tabular}{c|c|c|c|c|c|c|c|c|c|c|}
\hline
 $JD + \footnotesize{2400000}$ & Day & B\tiny{(mag)} & errB\tiny{(mag)} & V\tiny{(mag)} & errV\tiny{(mag)} & R\tiny{(mag)} & errR\tiny{(mag)} & I\tiny{(mag)} & errI\tiny{(mag)} & Telescope \\
\hline
56009.45  & 6.95 & 13.449 & 0.020 & 13.334 & 0.014 & 13.144 & 0.013 & 13.130 & 0.013 & AZT-8 \\
\hline
56010.48  &   7.98  &  - & -  & 13.332 & 0.026 &  13.156 & 0.017 &   13.078 & 0.017 & AZT-8 \\
\hline
56012.29  &   9.79  & - & - &  13.299 & 0.022 &  13.153 & 0.017  & 13.112 & 0.014 & LX200 \\
\hline
56018.41  &  15.91  & - & - & - & - &  13.096 &  0.021   & - & - & LX200 \\
\hline
56019.34  &  16.84   &  - & -  & - & - &  13.082 & 0.032   & - & - & LX200 \\
\hline
56021.35  &  18.85 & 13.611 & 0.044 & 13.166 & 0.038 & - & - & 12.911 & 0.032 & AZT-8 \\
\hline
56022.31  &  19.80   &  - & -  & 13.041 & 0.039 & 13.020 & 0.004   & - & - & LX200 \\
\hline
56023.38 &   20.84  & - & - &  13.201 & 0.031  & - & -  & - & - & LX200 \\
\hline
56027.38 & 24.88 & 13.964 & 0.027 & 13.323 & 0.025 & 13.011 & 0.023 & 12.895 & 0.021 & LX200 \\
\hline
56033.30 &   30.80  & - & - &  13.398 & 0.021 &  13.048 & 0.018  &  12.887 &  0.028 & LX200 \\
\hline
56039.39 &   36.89   & - & - &  13.440 & 0.042 &  13.075 & 0.029  & 12.823 & 0.028 & AZT-8 \\
\hline
56040.39  &  36.90   & - & - & - & - &  13.014 & 0.023  & 12.782 & 0.024 & LX200 \\
\hline
56041.25 &  38.75 & 14.319 & 0.019 & 13.355 & 0.016 & 13.022 & 0.028 & 12.754 & 0.017 & AZT-8 \\
\hline
56043.320  &  40.82  & - & -  & - & - &  13.087 & 0.038 & - & - & LX200 \\
\hline
56049.29 & 46.79 & 14.441 & 0.022 & 13.424 & 0.038 & - & - & 12.715 & 0.050 & AZT-8 \\
\hline
56050.41 &   47.91  & - & -  & 13.465 & 0.025 & 13.089 & 0.031 &   12.713 & 0.032 & LX200 \\
\hline
56063.36 & 60.86 & 14.674 & 0.022 & 13.431 & 0.010 & 12.970 & 0.014 & 12.659 & 0.011 & AZT-8 \\
\hline
56221.61 & 219.11 & 17.941 & 0.098 & 16.748 & 0.058 & 15.843 & 0.020 & 15.376 & 0.040 & AZT-8 \\
\hline
56235.64 &  233.14 &  - & -  & 16.840 & 0.110 & 15.894 &  0.029  & 15.411 & 0.027 & AZT-8 \\
\hline
56259.59 &  257.09 & - & -  & 17.093 & 0.053 &  16.058 &  0.032 &   15.620 & 0.036 & AZT-8 \\
\hline
56279.61 &  257.09  & - & -  & 17.240 & 0.152 &  16.293 &  0.086 &   15.804 & 0.036 & AZT-8 \\
\hline
56310.54 & 308.039 & - & -  & - & - & 16.520  & 0.133  & 16.015 &  0.064 & LX200 \\
\hline
56355.37 & 352.871 & - & -  & - & - & 16.783  & 0.123  &  16.613 & 0.089 & LX200 \\
\hline
 56362.39 & 359.891 & - & -  & - & - & 17.119  & 0.160  & 16.648 & 0.058 & LX200 \\
 \hline
 56364.40 & 361.898 & - & -  & - & - & -  & -  & 16.654  & 0.157  & LX200 \\  
 \hline
56373.45 & 370.949 & - & -  & - & - & 17.284  & 0.170  &  16.786 & 0.093  & LX200 \\
\hline
\end{tabular}
\end{table*}



\subsection{Supernova Parameters from Observations}
\label{observ_param}

  Photometry results for the SN 2012aw are presented in a number of works \citep{DallOra2014, Bose2013a, Munari2013,Bayless2013}. \editone{\cite{Bose2013a} estimate} the plateau duration $\approx 110$ days, Dall'Ora et al. \citep{DallOra2014} $\approx 100$ days. 
  
  
  \editone{We get} an absolute magnitude in the middle of the plateau in the $V$ band equal to $ -16.92$ mag, which is consistent with estimates from other works: $M_{v} = - 16.67 \pm 0.04$~mag \citep{Bose2013a}. 
 The luminosity peak at the early light curve of SN 2012aw in $U$, $B$, $V$, $R$, $I$ is reached at 8, 11, 15, 22, 24 days, respectively, thereby the supernova is similar to SN 1999em and SN 2004et \citep{Bose2013a}.

 	The observed photospheric velocities for SN 2012aw ( $v = 3.68 (\times 10^{3}) km/s$) were taken from \cite {Bose2013a}.
 	
 	\subsection{Estimation of observable parameters}
 	
 \editAn{We have estimated the observable parameters (the plateau duration $\Delta t$, the absolute magnitude $M_V$, and the photospheric velocity $u_{ph}$ at the middle of the plateau) from physical parameters (the explosion energy $E$, the mass of the envelope expelled $M$, and the pre-supernova radius $R$) using the relations found by  \citet{Litvinova1985}. Assuming $E = 2.0$~foe, $M = 25 M_\odot$, $R = 500 R_\odot $ corresponding to our model \mdlB we obtained: $\Delta t = 124$ days, $M_V = -17.5$, and $u_{ph} = 4.26\times 10^3$~km~s$^{-1}$. These results are not too different from the values described in the Section~\ref{observ_param}. }

 
 

 
 \section{Hydrodynamic model}
 \label{Hyd_mod}

\sneiip, like other type II supernovae, exhibit a wide variety of shapes of light curves.
The shape of the light curve is mainly influenced by such parameters as the mass of the ejected supernova envelope $M$, the radius of the pre-supernova $R$, the explosion energy $E$ and the chemical composition of the star \citep{Litvinova1985}.

We calculated a model which describes \editone{observational data} of the SN 2012aw, using the multi-energy group
radiation hydrodynamics code \stella \citep{Blinnikov2000, Blinnikov1998}.
 The advantage of the \stella is that it can simultaneously calculate hydrodynamics and energy transfer. The non-stationary transport equation \editone{is solved assuming LTE simultaneously with the hydrodynamic equations.}
 We calculated a grid of models in the parameter space  $M$, $R$, $^{56}Ni$, $E$ to search for a model describing observational data for the 2012aw supernova.

\begin{figure}
\begin{center}
  \includegraphics[width=\columnwidth,clip]{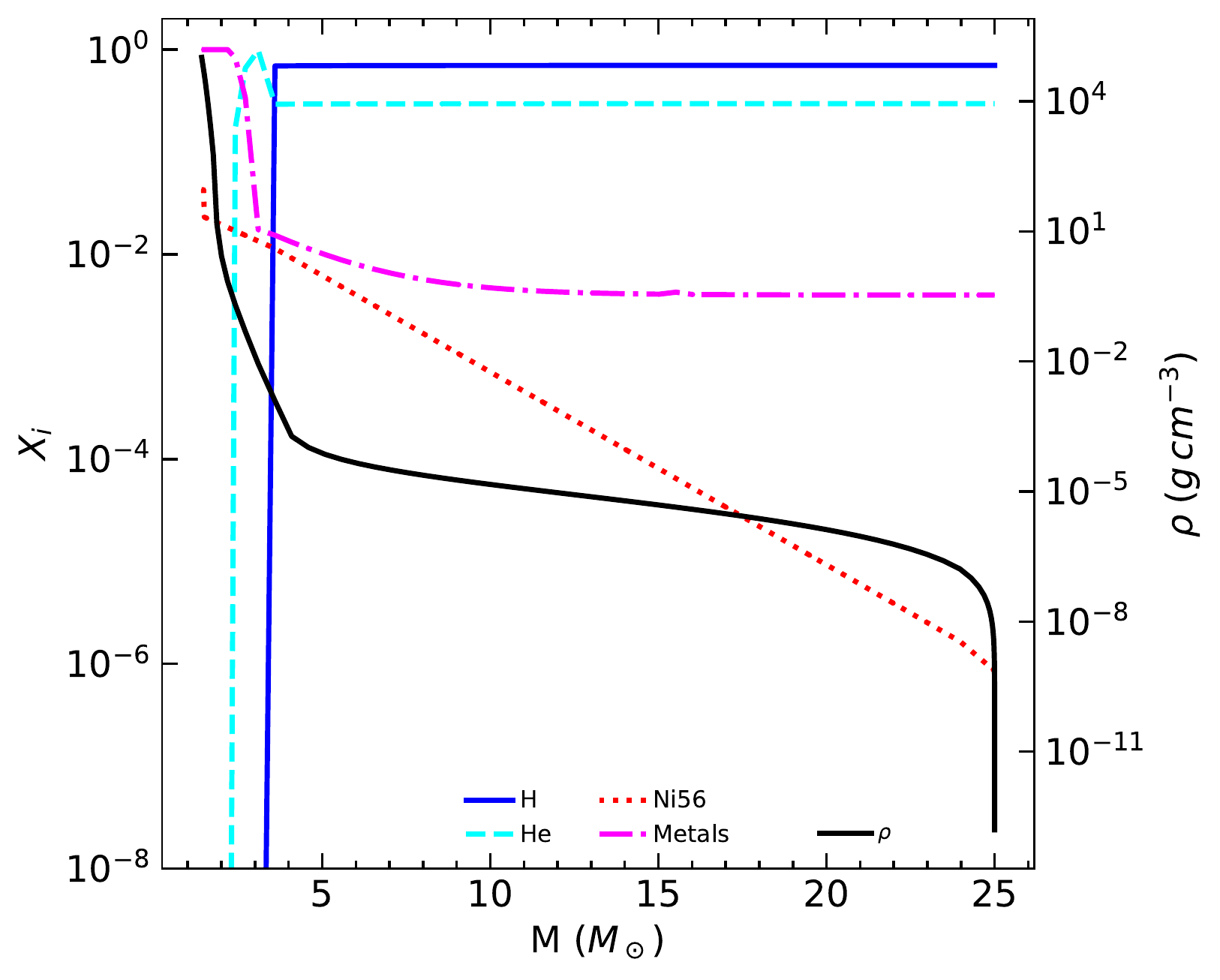}
   \caption{Distribution of chemical elements within the pre-supernova. \editone{The Y axis(right) displays the density, the Y axis(left) displays the relative content of the elements. Here "Metals" means the fraction of all elements heavier than He.}}
   
\label{figure6}
\end{center} 
\end{figure}

\editone{For pre-SN we use} a non-evolutionary polytropic model, like SN 1999em in the work of \cite{Baklanov2005}. 
Figure~\ref{figure6} \editone{shows} the \editone{density} distribution and \editone{the mass fraction} of chemical elements  {as a function of interior mass} within the pre-supernova. 
It is assumed power-law dependence of the temperature on the density \citep{Baklanov2005}.
\editone{In the center we isolate a dense core with the mass of 1.4 $M_\odot$, 
\editone{which collapses to a proto-neutron star}}.
 \editone{The explosion is initialized in \stella  as a thermal bomb just above the core mass}.
 
 The chemical composition of the host galaxy NGC~3351 is close to solar \citep{VanDyk2013, Fraser2012}, therefore, for the outer layers of the pre-supernova shell, \editone{we adopt mass fractions of hydrogen $\mathrm{X}=0.735$, helium $ \mathrm{Y}=0.248$, and the metallicity $\mathrm{Z}=0.17$.}	
 	
 The importance of joint fitting of light curves and expansion velocities of a supernova shell has been repeatedly emphasized in works with detailed modeling of supernovae \citep{Blinnikov2000, Baklanov2005, Utrobin2007}. 
\editone{This statement can be illustrated by fitting in two ways: first, the light curve only  and, second, the light curve in combination with photospheric velocities.} 	
 	
\begin{figure}
\begin{center}
   \includegraphics[width=\columnwidth,clip]{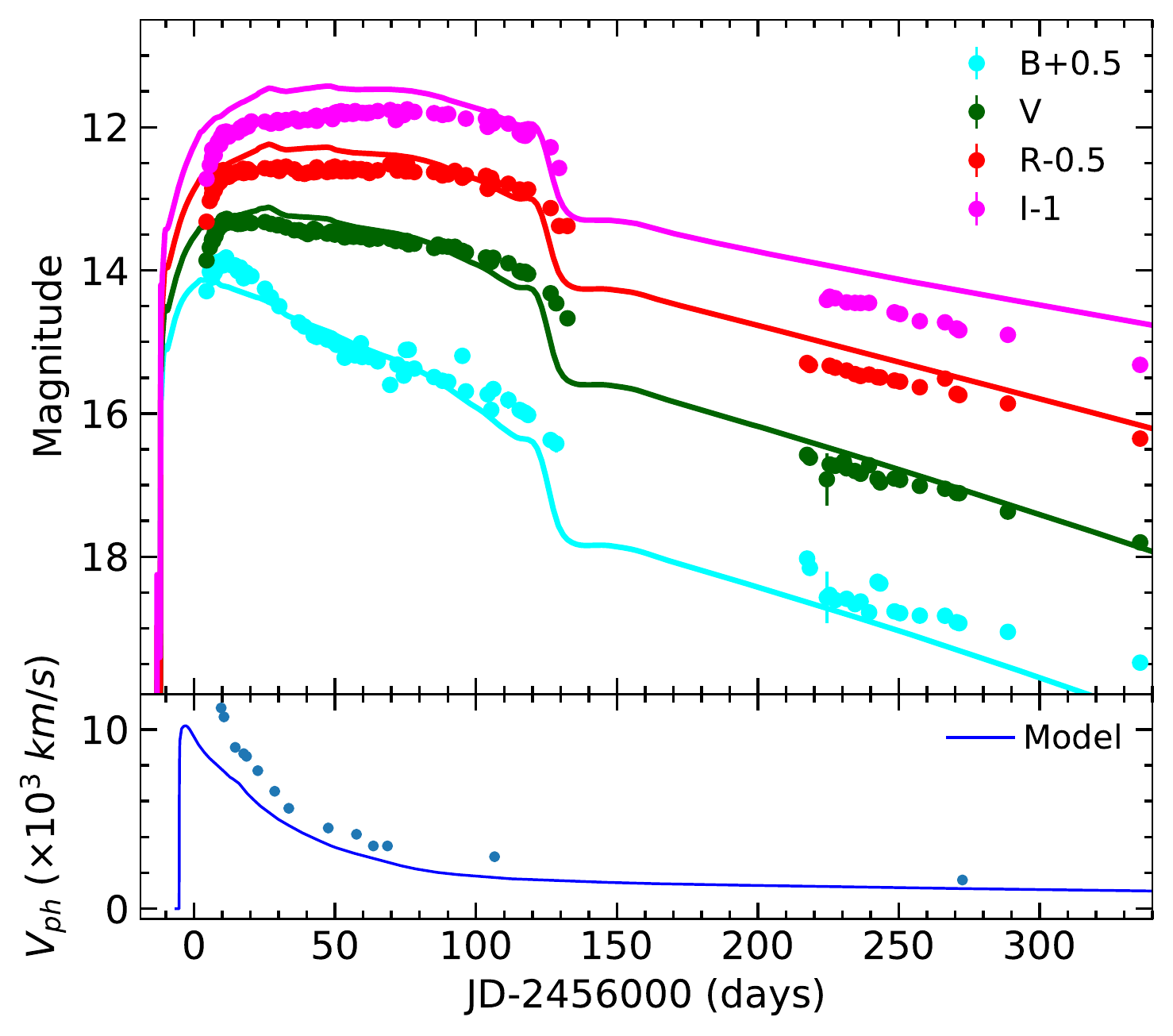}
   \caption{The model \mdlA, accounting for photospheric velocities. Dots indicate observational data, four light curve lines indicate calculated curve, blue line below indicate calculated photospheric velocities. Light curves show good agreement between observations and modeling, but photospheric velocities from observations and modeling do not match.}
\label{ubvvel_cat_R750_M25_Ni006_E20}
\end{center} 
\end{figure}

  \editone{In the first approach we come to the model \mdlA\footnote{{\mdlA: The model name contains the parameters that this model was calculated with.}} }
  shown in Figure~\ref{ubvvel_cat_R750_M25_Ni006_E20}. \editone{This model, however, demonstrates a poor agreement between the observed and calculated photospheric velocities}. It can be seen from the graph that the explosion energy in this model is not enough, and the shell scatters more slowly than was observed for SN 2012aw.

 	
 The observed photospheric velocities for SN 2012aw were taken from \cite {Bose2013a}. They were calculated from the absorption lines of Fe II in the late epochs and \editone{He I} in the early epochs after the supernova explosion.	
 	
\begin{figure}
\begin{center}
  \includegraphics[width=\columnwidth,clip]{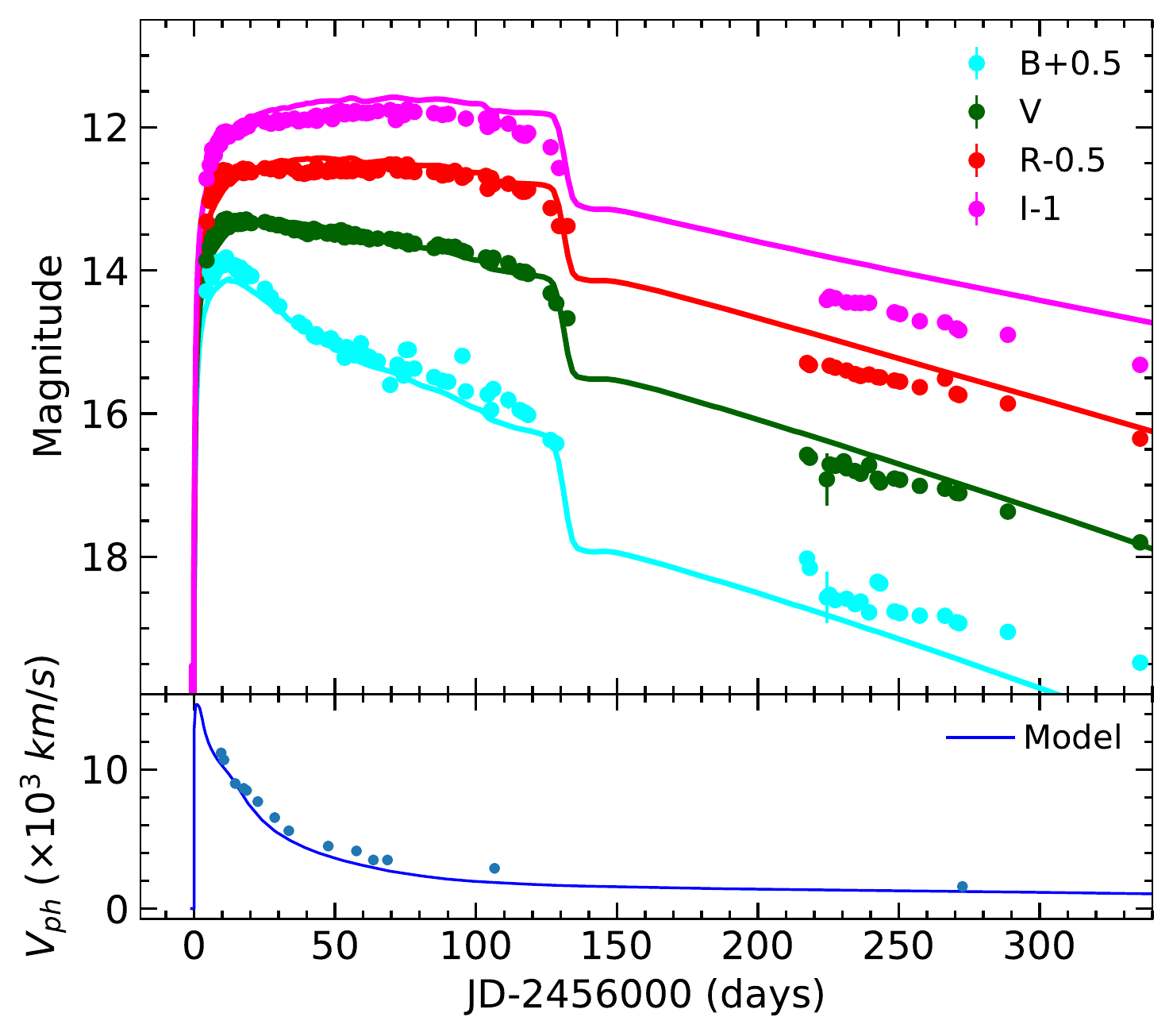}
   \caption{The best fit model when we are taking into account both light curves and photospheric velocities: \edittwo{\mdlB}}
\label{figure8}
\end{center} 
\end{figure}

 We applied the fitting procedure, which takes into account both the light curves and the photospheric velocities of the supernova. We selected the \edittwo{\mdlB} \editone{model, which provides a best fit to observational data of SN 2012aw} (Figure~\ref{figure8}) among other models.

 \editone{The pre-supernova radius in our model ($500 R_\odot$) is comparable to the 
value of $430 R_\odot$ reported by} \cite{DallOra2014}. The mass of the ejected shell ($M_{ej}=23.6 M_\odot, M_{tot}= 25 M_\odot$) in our model is larger than that estimated by \cite{DallOra2014} ($20 M_\odot$). \editone{Our estimate of $^{56}$Ni mass ($0.06 M_\odot$) is consistent with values 
 reported in previous works ($0.05 - 0.06 M_\odot$).}
 The mass of the $^{56}$Ni in our model is also equal to $0.06 M_\odot$. \editone{
  Previous estimates of SN 2012aw explosion energy --- 1 foe \citep{Bose2013a}, 
1.5 foe \citep{DallOra2014}, and 2 foe \citep{Bose2014} --- are \editAn{close}  
to our value of explosion energy (\editAn{2} foe).}

\section{Discussion}
 \label{discus}
 
     \editone{ A possible pre-supernova of the SN 2012aw was investigated by \citet{VanDyk2013}. They identified the object PTF12bvh as the progenitor of the SN 2012aw from archives of the Hubble Space Telescope, as well as from the ground-based observations in the near-infrared region. This field has been observed by the Hubble telescope at $F439W$, $F555W$ and $F814W$ between December 1994 and January 1995. \citet{VanDyk2013} estimated the magnitude of the pre-supernova as $V = 26.59$ mag.}
 	
\editone{ Investigating the nature of the pre-supernova, \citet{VanDyk2013} found that PTF12bvh is a red supergiant of class M3 with an effective temperature $ T_{\rm eff} =$  3600 K and a bolometric luminosity of $ M_{\rm bol} = - 8.29 $ mag [$ \log(L_{\rm bol}/L_\odot) = 5.21 \pm 0.03 $], effective radius $R_\sim 1040 \pm 100 R_\odot $. 
 The initial mass of the star was estimated as $ \sim 17 \div 18 M_\odot $. 
 Near the progenitor star, a significant amount of dust was noted.}

 \editone{The same object was identified as a pre-supernova in an earlier paper by \cite{Fraser2012}. 
 The bolometric luminosity was estimated as $ \log(L/ L_\odot) = 5.0 \div 5.6$, according to a mass of $ 14 - 26 M_\odot $. 
 The temperature estimate range is $3300 - 4500$ K, which gives a radius of $ R > 500 R_\odot $.} 
 
 \editone{Estimates of pre-supernova parameters of the SN 2012aw differ in different studies.
 Our task is to put together all the previous results, to compare them with our results and to analyze them.}
 
  \editone{Estimates of the radius of the SN 2012aw pre-supernova  star are as follows.
 \citet{DallOra2014} obtained the radius of the progenitor $\sim 430 R_\odot$ as a result of semi-analytical and hydrodynamic modelling. 
 Based on analytical relations, \citet{Bose2013a} obtained a not much different value   $\sim 337 \pm 67 R_\odot$. 
 A constraint on the radius of the pre-supernova star is also obtained from an analytical estimate $R > 500 R_\odot$ in \citep{Fraser2012}. 
According to \citet{VanDyk2013}, the pre-supernova had a radius $ R = 1040 \pm 100 R_\odot$. 
\citet{Martinez2019} derived  $R = 800 \pm 100 R_\odot $ from hydrodynamic modelling. 
We have got $R = 500 R_\odot$, which is close to the estimates found by \citet{DallOra2014} and \citet{Fraser2012}.}

\editone{The initial mass of \nifsx and the energy of the explosion do  agree much better in the estimates of different studies: 
$M(^{56}$Ni$) \sim 0.06 M_\odot$ \citep{DallOra2014}, \editAn{$M(^{56}$Ni$) \sim 0.06 M_\odot$ \citep{Hillier2019}}, $M(^{56}$Ni$)$ $\sim 0.058 \pm 0.002 M_\odot$ \citep{Bose2013a}, $M(^{56}$Ni$)$ $\sim 0.066 \pm 0.006 M_\odot$ \citep{Martinez2019}. 
Our estimate of $M(^{56}$Ni$) \sim 0.06 M_\odot$ is consistent with others.}

\editone{Explosion energy from other studies: $E \sim 1.5 \times 10^{51}$~erg \citep{DallOra2014}, $E \sim 1 \div 2 \times 10^{51}$ erg~\citep{Bose2013a}, $E \sim 2 \times 10^{51}$ erg \citep{Bose2014}, $E \sim 1.4 \times 10^{51}$ erg  \citep{Martinez2019}, \editAn{$E \sim 1.2 \times 10^{51}$~erg \citep{Hillier2019}.}
We have got the value $E = \editAn{2.0} \times 10^{51}$~erg.}
 
 \editone{The initial mass of the star is estimated from $12.5 \pm 1.5 M_\odot$ \citep{Fraser2016} to $21 M_\odot$  \citep{DallOra2014}. 
 \citet{VanDyk2013}  estimated the initial mass of a star in the range of $15-20 M_\odot$.} \editAn{\citet{Hillier2019} calculated the mass of progenitor as $15 M_\odot$. }
 
 \editone{The most important result of our model is on the mass of the envelope ejected during the explosion. 
 According to previous estimates we have rather high numbers for our object.
\cite{DallOra2014} rated the ejecta mass as $\sim 20 M_\odot$. 
\cite{Bose2013a} obtained a value of  $ 14 \pm 5 M_\odot $ with large error estimate.
\cite{Martinez2019} obtained $23_{-2}^{+1} M_\odot$. 
Our simulations with \stella have the most detailed physics in comparison with all cited papers and they yield the best pre-supernova mass $23.6 M_\odot$.
The latter number is appreciably higher than the upper limit $ 17 M_\odot$, 
which means the  `Red Supergiant Problem' problem persists \citep{Smartt2009AA, Davies2020, Kochanek2020}.
Moreover, recently there are more and more supernova models constructed for other objects with estimates of the ejecta mass
appreciably larger than the Smartt's limit: see, e.g. \cite{Utrobin2017}, \cite{Utrobin2019}.
Thus, our results give one more confirmation that the theory of pre-supernova evolution is not yet fully understood, and this question deserves further investigation.}

 \section{Conclusions}
 \label{concl}	

We report the results of our photometric observations of the \snaw
and compare it with the published data for this object.

To build our model we took into account both the light curves
and the photospheric velocities of SN 2012aw. This is an important
point that allows us to find the most suitable model among others.

We performed hydrodynamic modeling of both photometric and spectral data using the package
\stella and showed that the best agreement of the model with observations \editone{is found for the model} \edittwo{\mdlB}. 
\editone{ In this model the presupernova mass is  $25 M_\odot$ with the ejected $23.6 M_\odot$, the explosion energy is \editAn{2.0}~foe, the pre-supernova radius is  $500 R_\odot$, and the $^{56}$Ni mass is $0.06 M_\odot$.}
\editone{The total mass of  \snaw is higher by a factor of 1.5 compared with the upper Smartt's limit, which emphasizes the RSG Problem.}

\section*{Acknowledgements}

\editAn{The authors are very grateful to the referee, N.N.~Chugai, for his comments and valuable suggestions.}
The St. Petersburg University team acknowledges support from Russian Scientific Foundation grant 17-12-01029.
PB is sponsored by  grant  RFBR 21-52-12032 in his work on the \stella code development. 
SB is supported by grant RSF 19-12-00229 in his work on supernova simulations.

\section*{Data availability}
The photometric data are presented in Table ~\ref{table3}.
The details on calibrations are available from the first author upon request.

\bibliographystyle{mnras} 
\bibliography{arxiv_sn2012aw}

\vspace{0.5cm}\noindent


\bsp
\label{lastpage}
\end{document}